\documentclass[12pt]{article}

\usepackage{graphics,graphicx,fullpage,natbib,multirow}
\usepackage{amsmath,amssymb,verbatim,epsfig}
\usepackage[dvipsnames,usenames]{color}

\newtheorem{defin}{\bf Definition}

\def\ga{\mbox{Ga}}

\def\be{\mbox{Be}}

\def\bin{\mbox{Bin}}

\def\mul{\mbox{Mul}}

\def\no{\mbox{N}}

\def\un{\mbox{Un}}

\def\dir{\mbox{Dir}}

\def\E{\mbox{E}}
\def\V{\mbox{Var}}
\def\Cov{\mbox{Cov}}

\def\P{\mbox{P}}

\def\rest{\mbox{rest}}

\def\bx{{\bf x}}

\def\bz{{\bf z}}

\def\bI{\mbox{I}}

\def\bX{{\bf X}}

\def\bZ{{\bf Z}}
\def\bzero{{\bf 0}}

\newcommand{\balpha}{\boldsymbol{\alpha}}

\newcommand{\bepsilon}{\boldsymbol{\epsilon}}

\newcommand{\btheta}{\boldsymbol{\theta}}

\newcommand{\NB}{\mathbb{N}}

\begin{document}

\baselineskip=24pt

\title{\bf A binary factor model}
\author{{\sc Luis E. Nieto-Barajas} \\[2mm]
{\sl Department of Statistics, ITAM, Mexico} \\[2mm]
{\small {\tt luis.nieto@itam.mx}} \\}
\date{}
\maketitle

\begin{abstract}
The orthogonal factor model has been a very useful tool in uncovering covariance structures in a set of variables through a smaller set of underlying factors. This old model is suitable for continuous variables with unbounded support, since the most common assumption for the observables and the factors is multivariate normality. In this work, we propose a factor model for binary data. Factors are negative dependent, so they avoid each other. We study the theoretical properties of the model and carry out a full Bayesian inference. We illustrate the performance of our proposal with simulated and real data sets and compare with the traditional benchmark. 
\end{abstract}

\vspace{0.2in} \noindent {\sl Keywords}: Bayesian inference, dependent Bernoulli variables, dichotomous variables, latent variables.

\section{Introduction}
\label{sec:intro}

The objective of factor analysis is to describe the covariance relationship between several variables in terms of a few unobservable quantities called factors. This statistical technique was developed almost a century ago in psychometrics \citep{bartlett:37}. However, due to its popularity, it is still in use today \citep{beaducel&hilger:15}.

The factor model is defined as follows: Let $\bX'=(X_1,\ldots,X_p)$ be a vector of $p$ observable random variables of interest, let $\bZ'=(Z_1,\ldots,Z_q)$ be a vector of $q$ unobservable random factors, and let $\bepsilon'=(\epsilon_1,\ldots,\epsilon_p)$ be a vector of $p$ measurement errors such that 
\begin{equation}
\label{eq:fa}
\bX=\Lambda\bZ+\bepsilon,
\end{equation}
where $\Lambda=(\lambda_{j,k})$ is a matrix of parameters, usually called loadings, of dimension $p\times q$. 

Typical distributional assumptions on the model \eqref{eq:fa} are $\bZ\sim\no(\bzero,\bI)$, to define orthogonal factors, and $\bepsilon\sim\no(\bzero,\Psi)$ independent of $\bZ$, where $\bI$ is the identity matrix of dimension $q\times q$ and $\Psi=\mbox{diag}(\psi_1,\ldots,\psi_p)$ with $\psi_j$ are the specific variances for $j=1,\ldots,q$. These assumptions imply that the distribution of the observables is $\bX\sim\no(\bzero,\Sigma)$, where 
\begin{equation}
\label{eq:sigma}
\Sigma=\Lambda\Lambda'+\Psi.
\end{equation}
The factorization \eqref{eq:sigma} of the variance-covariance matrix $\Sigma$ of $\bX$ is the basis for the interpretation of the model.

Given the distributional assumptions of the orthogonal factor model, it is suitable for variables with unbounded support. In particular, if the variables of interest are dichotomous, the model \eqref{eq:fa} cannot be used. Typical solutions rely on the assumption of an underlying latent continuous response $\bX^*$ that generates binary outcomes by thresholding. Therefore, instead of computing Pearson correlations in factorization \eqref{eq:sigma}, tetrachoric correlations are used. Alternatively, replacing the latent response $\bX^*$ by $\bX$ in \eqref{eq:fa} and thresholding at zero, we obtain the probit regression model \citep{muthen:78} or the item response model in psychology \citep{takane&leeuw:87}. 

An alternative way to define the previous factor model is to consider conditional and marginal distributions of the form: 
\begin{equation}
\label{eq:cfa}
\bX\mid\bZ\sim\no(\Lambda\bZ,\Psi)\quad\mbox{and}\quad\bZ\sim\no(\bzero,\bI)
\end{equation}
The induced marginal distribution for $\bX$ is $\no(\bzero,\Sigma)$, with $\Sigma$ given in \eqref{eq:sigma}.

In this article, we propose an alternative factor analysis model for binary data, motivated by the conditional representation \eqref{eq:cfa} of the normal factor analysis model. Our factors jointly have a dirichlet distribution, which implies a marginal beta distribution for each factor. 

The contents of the rest of the paper is as follows: In Section \ref{sec:model} we define our model and study its properties. Section \ref{sec:inference} shows how to perform a full Bayesian analysis. The model is illustrated in Section \ref{sec:examples} and we finally conclude in Section \ref{sec:conclusion}. 

Before proceeding, we introduce the notation. Let $\bin(1,\theta)$ denote a Bernoulli distribution with probability of success $\theta\in[0,1]$. We use the binomial notation to avoid confusion with $\be(a,b)$, which denotes a beta distribution with parameters $a,b\in[0,\infty)$ and mean $a/(a+b)$. Let $\dir(\balpha)$ denote a dirichlet distribution with parameter vector $\balpha=(\alpha_1,\ldots,\alpha_q)$, $\alpha_j>0$, and $\mul(c,\btheta)$ denote a multinomial distribution with number of trials $c\in\NB$ and probability of success $\btheta=(\theta_1,\ldots,\theta_q)$ with $\theta_j\in[0,1]$ and $\sum_{j=1}^q\theta_j=1$.

\section{Model}
\label{sec:model}

Let $X$ and $Z$ be two random variables. We first note that if $X\mid Z\sim\bin(1,Z)$ and $Z\sim f_Z$ then the marginal distribution for $X$ is still Bernoulli with probability of success given by $\P(X=1)=\E_Z\{P(X=1\mid Z)\}=\E(Z)$. 

Let $\bX'=(X_1,\ldots,X_p)$ be a vector of $p$ binary responses and let $\bZ'=(Z_1,\ldots,Z_q)$ be a vector of $q<p$ latent factors. We assume that each response $X_j$ has a conditional distribution  
\begin{equation}
\label{eq:xz}
X_j\mid\bZ\sim\bin(1,\theta_j(\bZ))\quad\mbox{with}\quad
\theta_j(\bZ)=\E(X_j\mid\bZ)=\sum_{k=1}^q\omega_{jk}Z_k,
\end{equation}
where the loadings $\omega_{jk}\in[0,1]$ for $j=1,\ldots,p$, with $X_j$ and $X_l$ conditionally independent given $\bZ$ for $j\neq l$. The joint distribution of the latent factors is 
\begin{equation}
\label{eq:z}
\bZ\sim\dir(\balpha),
\end{equation}
where $\balpha=(\alpha_1,\ldots,\alpha_q)$ with $\alpha_k\geq 0$. 

Equations \eqref{eq:xz} and \eqref{eq:z} form a new factor model for dichotomous variables, equivalent to the normal factor model \eqref{eq:cfa}. The loadings $\omega_{jk}\in[0,1]$ are constrained to the unit interval and represent the importance of variable $j$ in the factor $k$. No further constraints are needed to ensure that $\theta(\bZ)\in[0,1]$: if $\omega_{jk}=1$ for all $k$ then $\theta_j(\bZ)=1$; and if $\omega_{jk}=0$ for all $k$ then $\theta(\bZ)=0$, with probability one, for $j=1,\ldots,p$. 

The new factors $\bZ$ are continuous variables in the $q$-dimensional simplex and are therefore negatively correlated. In fact, the variance-covariance matrix $\V(\bZ)$, has diagonal elements $\V(Z_k)=\widetilde\alpha_k(1-\widetilde\alpha_k)/(\alpha_0+1)$ and off-diagonal elements $\Cov(Z_k,Z_h)=-\widetilde\alpha_k\widetilde\alpha_h/(\alpha_0+1)$, where $\widetilde\alpha_k=\E(Z_k)=\alpha_k/\alpha_0$ and $\alpha_0=\sum_{k=1}^q\alpha_k$. The latter is sometimes called total mass and is interpreted as precision (reciprocal of variance). The larger $\alpha_0$ implies a smaller variance for the factors, whereas the smaller $\alpha_0$ means larger variance. The total variance encoded in the factors is ${trace}\{\V(\bZ)\}=\sum_{k=1}^q\V(Z_k)$, so the proportion of each factor contribution can be easily calculated by the ratio 
\begin{equation}
\label{eq:pTV}
\frac{\V(Z_k)}{{trace}\{\V(\bZ)\}}=\frac{\widetilde\alpha_k\left(1-\widetilde\alpha_k\right)}{\sum_{h=1}^q\widetilde\alpha_h\left(1-\widetilde\alpha_h\right)}.    
\end{equation}

As mentioned above, the induced marginal distribution for each response $X_j$ is Bernoulli with probability of success 
\begin{equation}
\label{eq:meanx}
\E(X_j)=\E\E(X_j\mid\bZ)=\E\left\{\sum_{k=1}^q\omega_{jk}Z_k\right\}=\sum_{k=1}^q\omega_{jk}\widetilde\alpha_k.
\end{equation}
Moreover, the responses $X_j$, $j=1,\ldots,p$ can be positive or negative correlated. In fact, for any pair $(X_j,X_l)$ with $j\neq l$,
$$\Cov(X_j,X_l)=\E\Cov(X_j,X_l\mid\bZ)+\Cov\left\{E(X_j\mid\bZ),E(X_l\mid\bZ)\right\}.$$
Due to conditional independence between $X_j$'s, the covariance simplifies to 
$$\Cov(X_j,X_l)=\sum_{k=1}^q\sum_{h=1}^q\omega_{jk}\omega_{lh}\Cov(Z_k,Z_h).$$
Finally we get
\begin{equation}
\label{eq:cov}
\Cov(X_j,X_l)=\sum_{k=1}^q\omega_{jk}\omega_{lk}\frac{\widetilde\alpha_k\left(1-\widetilde\alpha_k\right)}{\alpha_0+1}-\sum_{k\neq h}^q\omega_{jk}\omega_{lh}\frac{\widetilde\alpha_k\widetilde\alpha_h}{\alpha_0+1}.
\end{equation}

Note that both expressions \eqref{eq:meanx} and \eqref{eq:cov}, are functions of the marginal mean of the factors $\widetilde\alpha_k$ and show an inverse relation with the total mass $\alpha_0$. Larger values of $\alpha_0$ cause the covariance between $X_j$ and $X_l$ to be more concentrated around zero (less variance), whereas smaller values of $\alpha_0$, produce disperser or extremer values, positive or negative, i.e. larger variance. 

The model parameters are the loading matrix $\Omega=(\omega_{jk})$ for $j=1,\ldots,p$, and $k=1,\ldots,q$; and the factor parameters $\balpha'=(\alpha_k)$, for $k=1,\ldots,q$. In order to estimate them, we suggest following a Bayesian approach that will be described in the following section.

\section{Bayesian inference}
\label{sec:inference}

Let $\bX_1,\ldots,\bX_n$ be a sample of size $n$ from the model given by equations \eqref{eq:xz} and \eqref{eq:z}. Let us assume for the moment that together with each observation $\bX_i$ we have available the latent factor scores $\bZ_i$ for $i=1,\ldots,n$. In this case, the extended likelihood for $(\Omega,\balpha)$ is $$f(\bx,\bz\mid\Omega,\balpha)=f(\bx\mid\bz,\Omega)f(\bz\mid\balpha),$$ where 
$$f(\bx\mid\bz,\Omega)=\prod_{i=1}^n\left\{\prod_{j=1}^p\theta_j(\bz_i)^{x_{ij}}\left(1-\theta_j(\bz_i)\right)^{1-x_{ij}}\right\}$$
with $\theta_j(\bz_i)=\sum_{k=1}^q\omega_{jk}z_{ik}$ and 
$$f(\bz\mid\balpha)=\prod_{i=1}^n\left\{\frac{\Gamma(\sum_{k=1}^q\alpha_k)}{\prod_{k=1}^q\Gamma(\alpha_k)}\prod_{k=1}^q z_{ik}^{\alpha_k-1}I\left(\sum_{k=1}^q z_{ik}=1\right)\right\}.$$

To perform a Bayesian inference, we express our prior knowledge on $(\Omega,\balpha)$. We assume that $\omega_{jk}\sim\be(a_\omega,b_\omega)$ independently for all $j$ and $k$, together with $\alpha_k\sim\ga(1,c_\alpha)$ independently for all $k$, for given $a_\omega>0$, $b_\omega>0$ and $c_\alpha>0$. 

The posterior distribution for $\Omega$ and $\balpha$ will be characterised through their full conditional distributions. These are:

\begin{itemize}
\item[(i)] Conditional distribution for $\omega_{jk}$, $j=1,\ldots,p$, $k=1,\ldots,q$
$$f(\omega_{jk}\mid\bx,\bz,\rest)\propto\left[\prod_{i=1}^n\theta_j(\bz_i)^{x_{ij}}\left\{1-\theta_j(\bz_i)\right\}^{1-x_{ij}}\right]\omega_{jk}^{a_\omega-1}(1-\omega_{jk})^{b_\omega-1}I_{[0,1]}(\omega_{jk}).$$
\item[(ii)] Conditional distribution for $\alpha_k$, $k=1,\ldots,q$
$$f(\alpha_k\mid\bz,\rest)\propto \frac{\Gamma^n(\sum_{j=1}^q\alpha_j)}{\Gamma^n(\alpha_k)}\left(e^{-c_\alpha}\prod_{i=1}^nz_{ik}\right)^{\alpha_k}I_{[0,\infty)}(\alpha_k).$$
\end{itemize}
Both previous conditional distributions depend on the latent factor scores $\bz$, however, they are never observed. We can extend our set of full conditional distributions to include the posterior conditional distribution of the factors, which allows us to predict the factor scores. 
\begin{itemize}
\item[(iii)] Conditional distribution for $z_{ik}$, $i=1,\ldots,n$, $k=1,\ldots,q-1$
$$f(z_{ik}\mid\bx,\rest)\propto\left[\prod_{j=1}^p\theta_j(\bz_i)^{x_{ij}}\left\{1-\theta_j(\bz_i)\right\}^{1-x_{ij}}\right]z_{ik}^{\alpha_k-1}z_{iq}^{\alpha_q-1}I\left(z_{ik}\leq 1-\sum_{j\neq k}^{q-1}z_{ij}\right),$$
where $z_{iq}=1-\sum_{j=1}^{q-1}z_{ij}$.
\end{itemize}

With the previous conditional distributions (i)--(iii) we can implement a Gibbs sampler. However, none of them is of standard form, so we will require to include Metropolis-Hastings steps \citep{tierney:94}. We suggest using random walks for each parameter/factor based on uniform distributions around the current value of the chain. Specifically, for iteration $r+1$ we sample: $\omega_{j,k}\sim\un(\omega_{jk}^{(r)}-\delta_\omega,\omega_{jk}^{(r)}+\delta_\omega)$; $\alpha_k\sim\un(\alpha_{k}^{(r)}-\delta_\alpha,\alpha_{k}^{(r)}+\delta_\alpha)$; and $z_{ik}\sim\un(z_{ik}^{(r)}-\delta_z,z_{ik}^{(r)}+\delta_z)$, where $\delta_\omega$, $\delta_\alpha$ and $\delta_z$ are tuning parameters. The limits of the uniforms are constrained to lie within the conditional supports. We accept the proposals with the ratio of the conditional distributions evaluated at the new simulated value and the current value. 

The tuning parameters are adapted to achieve an optimal acceptance rate in the interval $[0.3,0.4]$. Following \cite{roberts&rosenthal:09}, we use batches of $100$ iterations, and for each batch $l$, we compute the average acceptance rate for each set of parameters/factors $\{\omega_{jk}\}$, $\{\alpha_k\}$, and $\{z_{ik}\}$, say $AR^{(l)}$, dropping the subindex. Then increase $\delta^{(l+1)}=\delta^{(j)}1.5^{1/\sqrt{l}}$ if $AR^{(l)}>0.4$; and decrease $\delta^{(l+1)}=\delta^{(l)}1.5^{-1/\sqrt{l}}$ if $AR^{(l)}<0.3$. We take $\delta_\omega^{(0)}=\delta_z^{(0)}=0.25$ and $\delta_\alpha^{(0)}=1$ as starting values.

\section{Rotations}
\label{sec:rotations}

It is well known that the parameters of the orthogonal factor model \eqref{eq:cfa} are not entirely identifiable. To see this, we consider an orthogonal matrix $T$, such that $T\,T'=T'T=\bI$, then the conditional expectation of $\bX$ can be written as $$\E(\bX\mid\bZ)=\Lambda\bZ=\Lambda\,T\,T'\,\bZ=\Lambda^*\bZ^*,$$
where $\Lambda^*=\Lambda\,T$ is a new loading matrix and $\bZ^*=T'\,\bZ$ is a new factor vector. Moreover, $\E(\bZ^*)=T'\,\E(\bZ)=\bzero$ and $\V(\bZ^*)=T'\,\bI\,T=\bI$, therefore, $\bZ^*\sim\no(\bzero,\bI)$. The new $(\Lambda^*,\bZ^*)$ are named \textit{rotations} of the original $(\Lambda,\bZ)$ and satisfy the same equations of model \eqref{eq:cfa}. 

Our binary model of Section \ref{sec:model} has conditional expectation, which can be written in matrix notation as $\E(\bX\mid\bZ)=\Omega\bZ$, where each element of the vector is given in \eqref{eq:xz}. By proceeding as above, we can consider an orthogonal matrix $T$ such that $$\E(\bX\mid\bZ)=\Omega\bZ=\Omega\,T\,T'\bZ=\Omega^*\bZ^*,$$
where $\Omega^*=\Omega\,T$ is a new loading matrix and $\bZ^*=T'\,\bZ$ is a new factor vector. 

However, in this case, since the distribution of the original factors $\bZ$ is dirichlet, the rotation $\bZ^*$ does not preserve the dirichlet distribution. To see this, consider the case $q=2$, where the general expression for a rotation matrix with angle $\pi$ is 
$$T=\left(\begin{array}{cc} 
\cos\pi & -\sin\pi \\
\sin\pi & \cos\pi
\end{array}
\right).$$
Then, $\bZ^*$ becomes 
$$\bZ^*=\left(\begin{array}{c}
Z_1\cos\pi-Z_2\sin\pi\\
Z_1\sin\pi+Z_2\cos\pi
\end{array}\right),$$
which does not satisfy $Z_1^*+Z_2^*=1$. In any case, we can always compute the variance of the new factor vector as $\V(\bZ^*)=T'\,\V(\bZ)\,T$. 

Our Bayesian inference procedure, outlined in Section \ref{sec:inference}, is based on proper prior distributions, so as long as these priors are not too vague and there are enough data, the posterior distribution is proper. The MCMC sampler will provide a unique estimate for $\Omega$ up to a permutation. 

Although not required, but if the configuration of the factor loadings is not appealing for interpretation, we suggest rotating the point estimate $\widehat\Omega$ with one of the most common rotation techniques such as \textit{varimax} \citep{kaiser:58} for an orthogonal rotation or \textit{promax} \citep{hendrickson&white:64} for an oblique rotation. Both of these rotation methods are included in the \cite{r-package:26}. 

After rotating the loading matrix $\Omega$, we use the same rotation matrix, say $T$, to rotate the scores $\bZ=(z_{ik})$ for $i=1,\ldots,n$ and $k=1,\ldots,k$, such that the new matrix of scores is $\bZ^*=\bZ\,T$. 

\section{Illustrations}
\label{sec:examples}

\subsection{Simulation study}

We first test our model and inference procedure with a controlled scenario. We take $p=7$ observable binary variables and $q=3$ latent factors. We define the factor parameters $\alpha_k=1/2$ for $k=1,\ldots,q$ and a loading matrix with values given by the first three columns in Table \ref{tab:sim}.

As is customary in factor analysis \citep[e.g.][]{johnson&wichern:02}, we added a box to the largest number row-wise that corresponds to the factor $k$ that better represents the variable $j$. The numbers were chosen so that each variable is clearly represented by one of the factors. 

We took a sample of size $n=500$ using, as generating model, the one described in Section \ref{sec:model}. We fit these data using our inferential procedure in Section \ref{sec:inference} with prior distributions defined with $a_\omega=b_\omega=1/2$ and $c_\alpha=1$. The proposed adaptive algorithm works pretty well, achieving the target acceptance rate by the fifth batch. An initial assessment suggested that the factor parameters $\alpha_k$ have chains that are highly correlated. Therefore, we run the Gibbs sampler for $100,000$ iterations with a burn-in of $10,000$ and a thinning of $100$. 

We assess the fit of the model by computing the logarithm of the pseudo marginal likelihood (LPML) proposed by \cite{geisser&eddy:79} and defined as $LPML=\sum_{i=1}^n\log CPO_i$, where the conditional predictive ordinates are given by $CPO_i=f(\bx_i\mid\bx_{-i})$. These ordinates can be easily approximated via Monte Carlo. Larger values of $LPML$ indicate a better fit. 

Since the distributions involved in the definition of our binary factor model \eqref{eq:xz}-\eqref{eq:z}, as well as the prior distributions, are of standard form, we could use generic Bayesian inference packages like JAGS \citep{plummer:23}, however the running time is very large, of around two hours for each run, therefore we decided to implement the algorithm described in Section \ref{sec:inference} in Fortran. 

We played with a set of values for $q\in\{2,3,4\}$ and the fit statistics together with the running times, in parentheses, are: for $q=2$, $LPML=-2151$ ($6.06$ mins.); for $q=3$, $LPML=-2088$ ($11.96$ mins.); and for $q=4$, $LPML=-2094$ ($17.25$ mins.). We note that for every additional factor dimension, the running times increase on average in 6 minutes, for this sample size. Moreover, our fit statistic is able to identify the correct number of factors, with the largest LPML value for $q=3$. 

Considering $q=3$, the estimated values of the factor parameters are as follows. The posterior mean and a $95\%$ credible interval (CI) are reported: $\widehat\alpha_1=0.43$ and $[0.31,0.61]$; $\widehat\alpha_2=0.39$ and $[0.30,0.52]$; and $\widehat\alpha_1=0.46$ and $[0.28,0.70]$. All CI's contain the true values. 

Finally, we present estimates (posterior means) of the loading matrix $\Omega$. We show estimates $\widehat\Omega_1$ and $\widehat\Omega_2$ for two different chains. The numbers are included in columns 3--6 and 7--9, respectively, of Table \ref{tab:sim}. There is complete agreement between the two matrix estimates, with the only difference that the columns are permuted. Although the point estimates are not identical to the true values, their $95\%$ CI (not shown) do contain the true values. The important thing is that the interpretation of the loadings is kept unaffected with respect to the original setting. 

\subsection{Real data analysis}

The Mexico Ministry of Health, through its Epidemiology Direction, maintains a record of all respiratory infections. The original database contains all 2026 registered cases and can be found at {https://www.gob.mx/salud/documentos/datos-abiertos-152127}. We filter the data and only kept the confirmed cases of COVID19. 

The resulting data set contains $n=1,814$ cases and $p=12$ explanatory variables, which are binary indicators of the patients: $X_1=$ women, $X_2=$ hospitalized, $X_3=$ pneumonia, $X_4=$ adult ($\geq 50$ years old), $X_5=$ diabetes, $X_6=$ epoc, $X_7=$ asma, $X_8=$ inmunosupressed, $X_9=$ hypertension or cardiac disease, $X_{10}=$ obesity, $X_{11}=$ chronic kidney failure, $X_{12}=$ smoker. 

We compute the indicators proportions (averages) for the variables in the dataset and report them in Table \ref{tab:eda}. There are slightly more women than men, almost half of the cases were hospitalized, and one third of the cases were adults. Six of the twelve indicators have percentages with values less than $10\%$. 

We fit our model and run the Gibbs sampler for $120,000$ iterations with a burn-in of $30,000$ and a thinning of $100$. The prior distributions were the same as for the simulation study. We also played with $q\in\{2,3,4\}$ factors. The fit statistic and the running times are: for $q=2$, $LPML=-7377$ ($26$ mins.); for $q=3$, $LPML=-6955$ ($50$ mins.); and for $q=4$, $LPML=-6936$ ($84$ mins.). 

Table \ref{tab:real} shows all posterior inferences for $\Omega$ and $\balpha$ for the three values of $q$. For $q=2$ (second and third columns), we get the worst fitting according to the LPML, the first factor is represented by the single variable WOMEN and the second factor by the rest of the variables, except ASMA which is equally unrepresented by the two factors. For $q=3$ (fourth to sixth columns), the loading rotated with varimax shows that one of the factors (second) is represented by WOMEN, other factor (first) by HOSP and PNEUM and the third factor by ADULT, DIABET, EPOC, CARDIO, OBESITY, KIDNEY; we also note that three variables, ASMA, INMUNO and SMOKE are not represented by any of the three factors. 

For $q=4$ (seventh to tenth columns), although the LPML favours these number of factors, one of them, after rotation with varimax, does not represent any of the variables. This suggests that $q=3$ is the best configuration. 

The last three rows in Table \ref{tab:real} show the posterior estimates of $\alpha_k$. We first analyse the total precision $\alpha_0$ achieved for the three values of $q$: for $q=2$, $\alpha_0=1.48$; for $q=3$, $\alpha_0=1.09$; and for $q=4$, $\alpha_0=1.06$. This means that with $q=4$ the model explains a little more variance than with $q=3$, and with $q=2$ the model explains the least variance of the three. Therefore, in addition to the LPML fit statistic, the total precision $\alpha_0$ is another indicator of how good a model is. 

Further interpreting the factors with $q=3$, we can say that the first factor puts together HOSP and PNEUM, which makes sense since pneumonia is a severe disease that usually requires hospitalization, we could name this factor \textit{covid complications}. The third factor gathers other diseases, like DIABET, EPOC, CARDIO, OBESITY and KIDNEY that usually appear as an ADULT, we could name this factor \textit{adult comorbidities}. Finally, the second factor is only represented by WOMEN, which says that the \textit{gender} has nothing to do with the \textit{covid complications} or \textit{adult diseases}.

In addition, we take $q=3$ to determine the importance of each factor. Since the reported loading matrix has been rotated, we also rotate the variance-covariance matrix of the original factors to produce $\V(\bZ^*)$. Using the same reasoning as in \eqref{eq:pTV}, the most important factors are the first (\textit{covid complications}) and the third (\textit{adult comorbidities}) each accounting for $36\%$ of the variance, while the least important is the second factor (\textit{gender}) with $28\%$ of the variance. The $95\%$ CI for $\alpha_k$'s do not intersect, confirming that the importance of the unrotated factors is distinctive to each other. 

\begin{figure}
\centering
\includegraphics[scale=0.3]{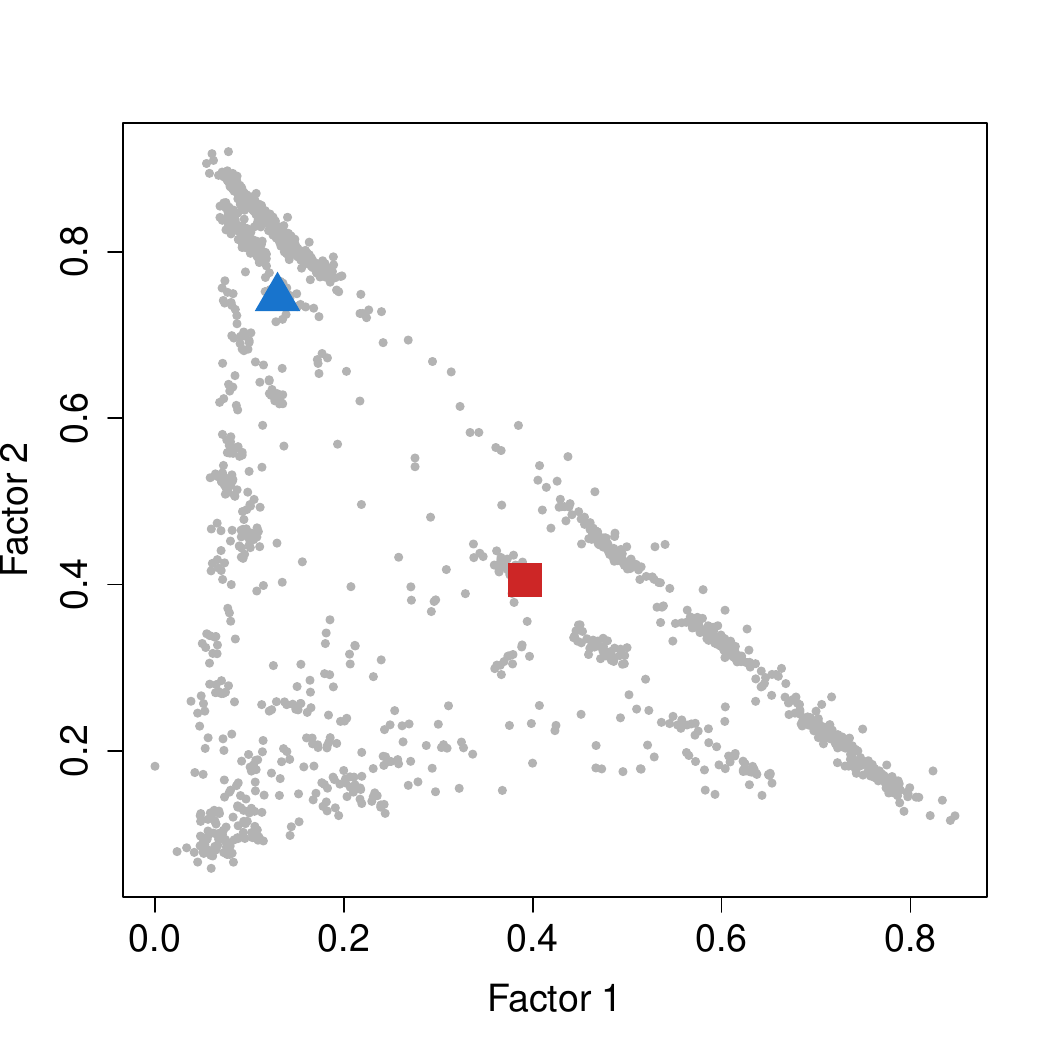}
\includegraphics[scale=0.3]{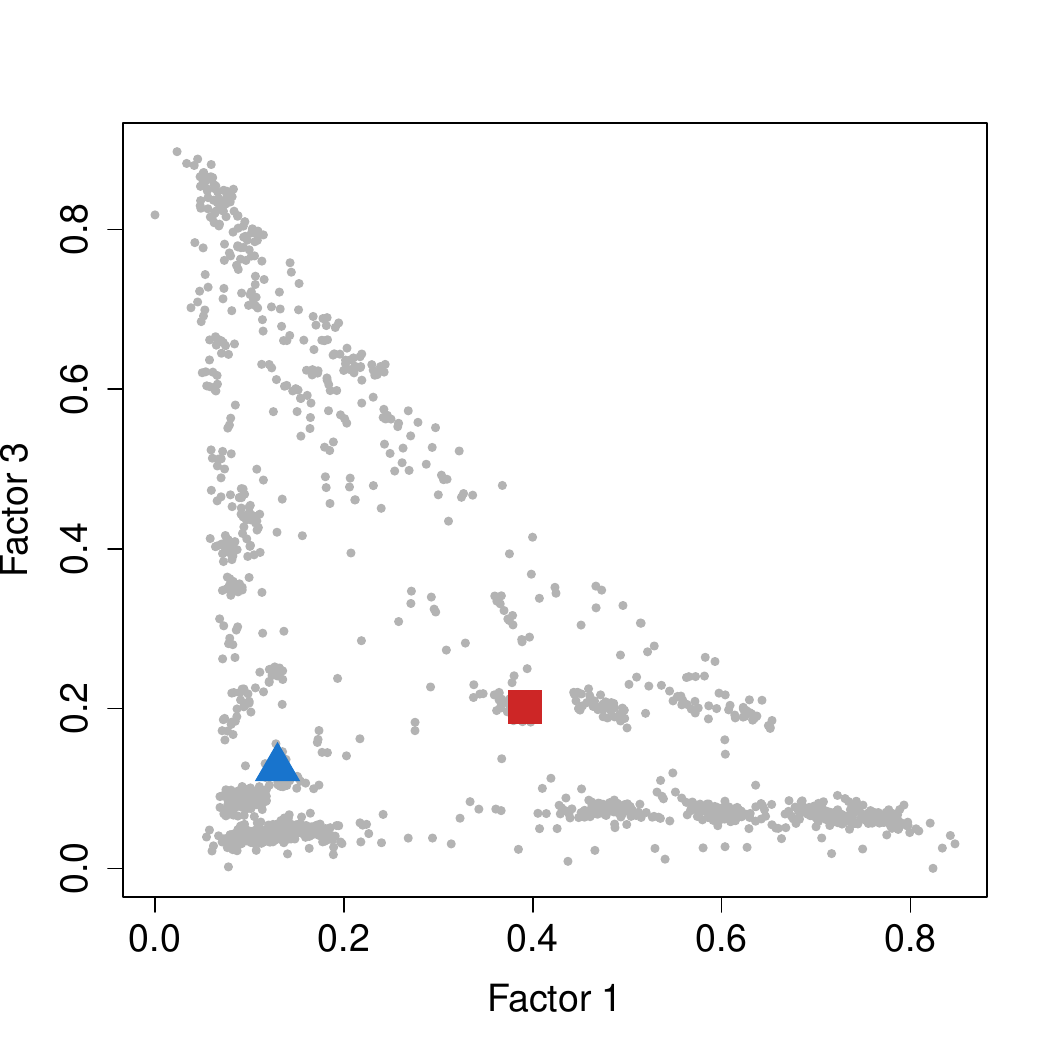}
\includegraphics[scale=0.3]{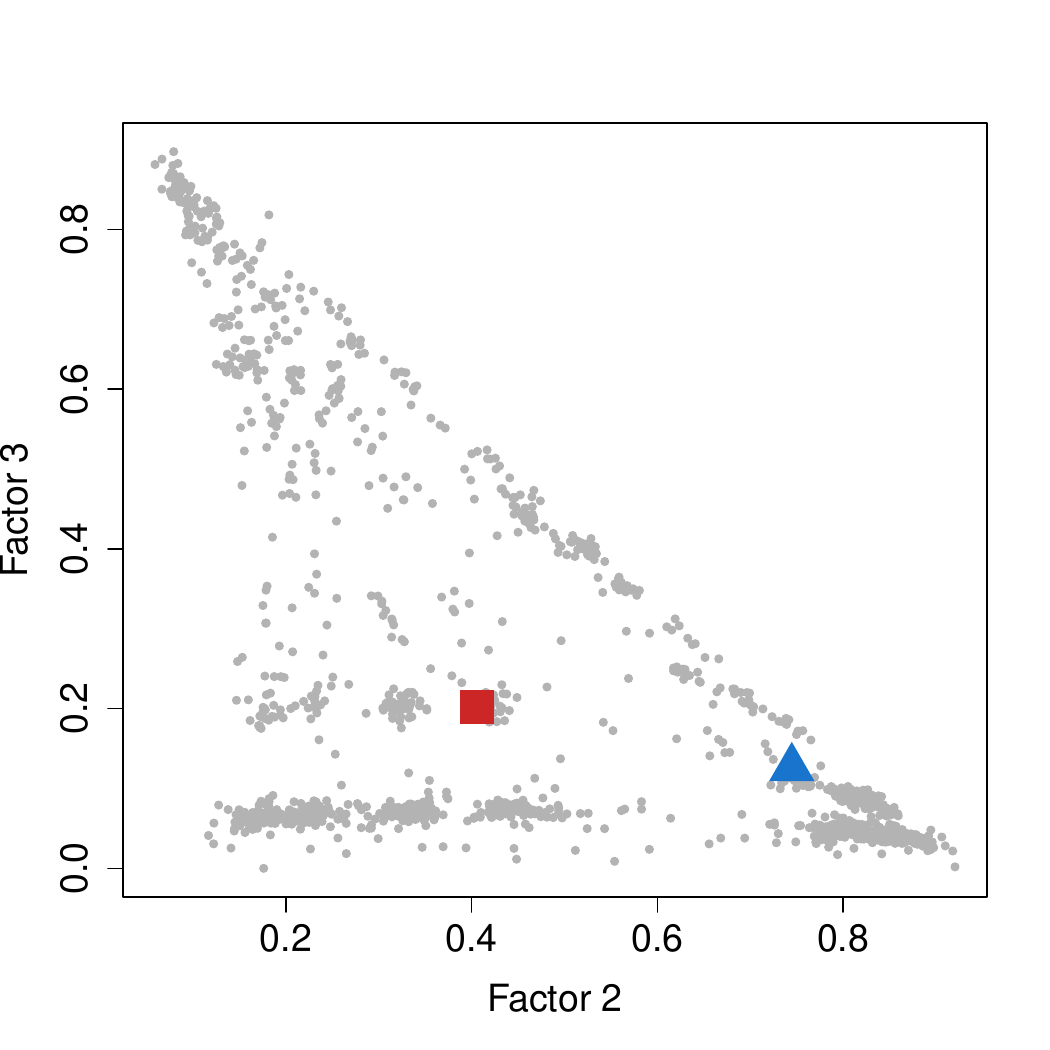}
\includegraphics[scale=0.3]{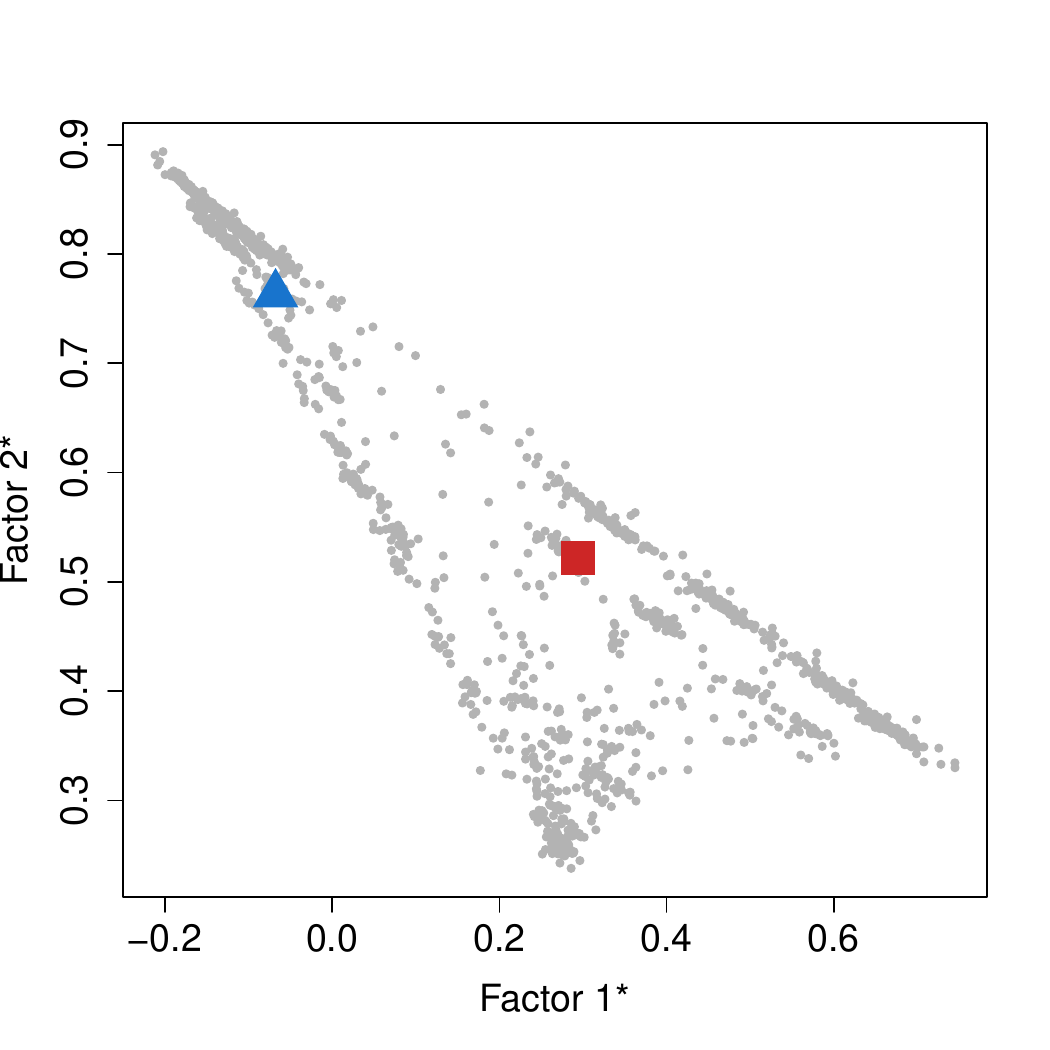}
\includegraphics[scale=0.3]{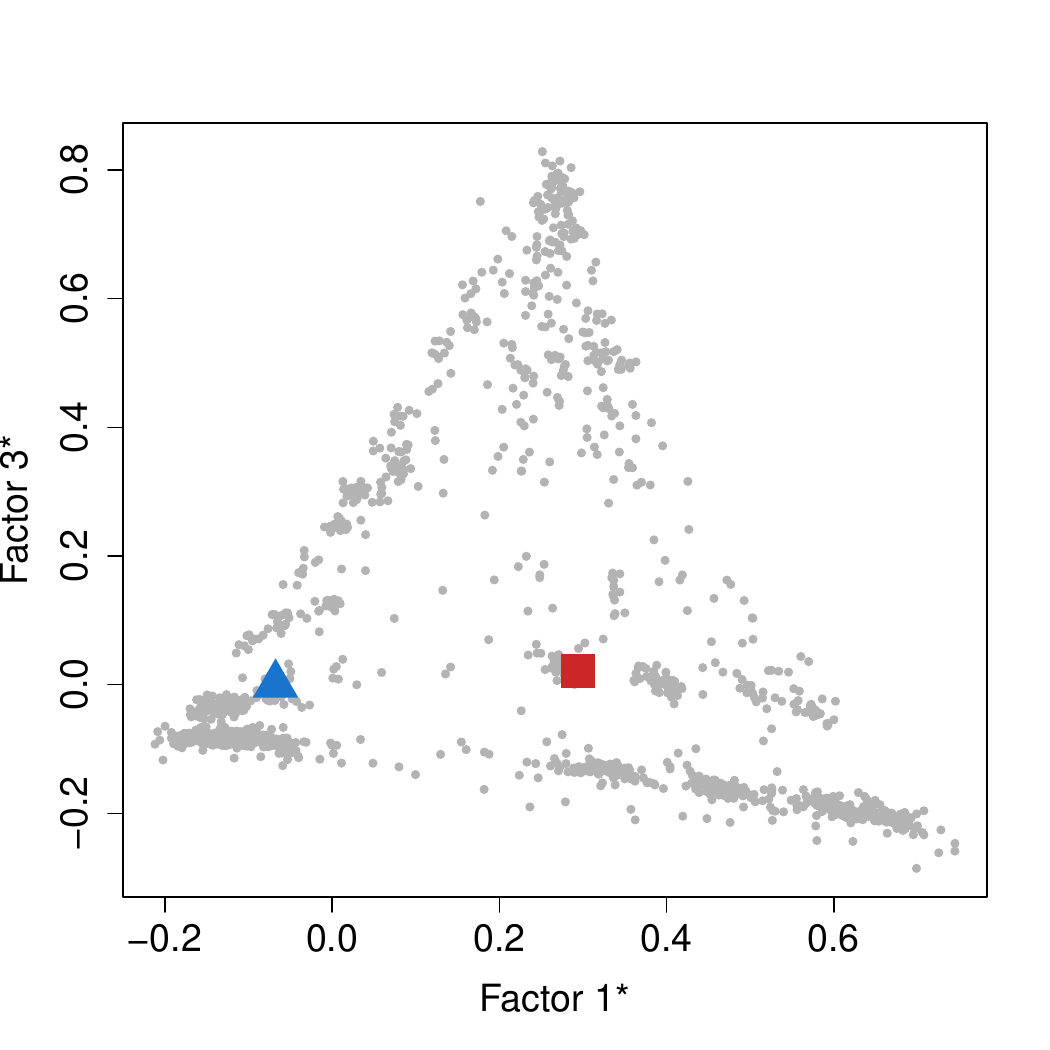}
\includegraphics[scale=0.3]{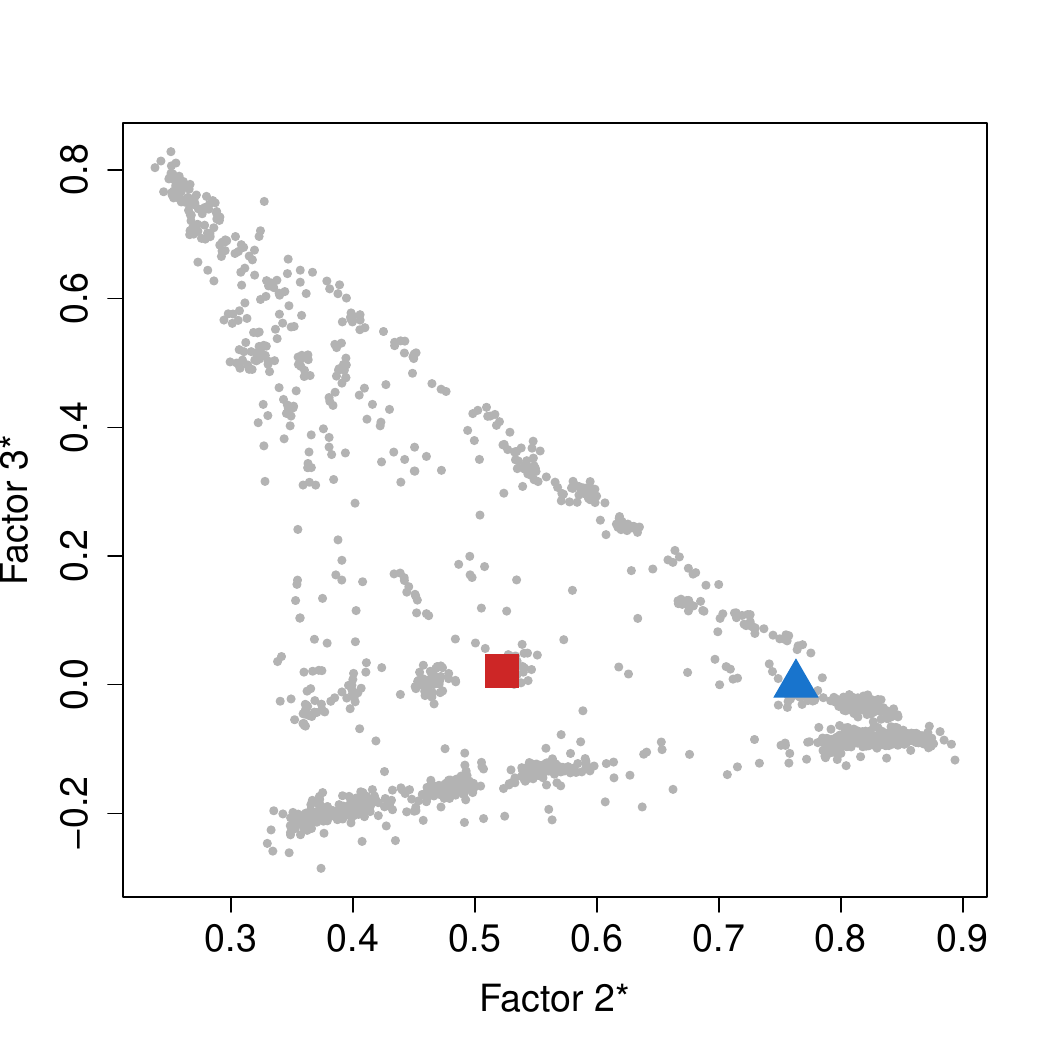}
\caption{COVID19 data. Score plots for original factors (top) and rotated factors (bottom). Red square and blue triangle are two specific individuals.}
\label{fig:scores}
\end{figure}

As a by product of our Bayesian inferential procedure, we can also produce dispersion diagrams of the individuals scores in the factor space. To show the impact of rotation, we report in Figure \ref{fig:scores} the original (top row) and rotated scores (bottom row). For the original scores, we see the sum one constraint, with all data points in the simplex. After rotation, the dispersion diagrams are clearly deformed away from the simplex. In all diagrams, we highlight two points: the red square corresponds to a hospitalized adult woman; the blue triangle correspons to a hospitalized adult man with diabetes, epoc, cardiac complications and kidney failure. 

\subsection{Model comparison}

In this section, we reanalyse our COVID19 dataset with the traditional procedures. We first compute the tetrachoric coefficients \citep{bonett&price:05} to produce a correlation matrix and use the normal factor model \eqref{eq:cfa} in R \citep{r-package:26} using the command \texttt{factanal} with $q=3$ factors and default \textit{varimax} rotation. Loadings estimates are shown in Table \ref{tab:tetra}. 

We see that the new three factors formed are somehow different from the ones we got with our model. The new first factor is compared to our third factor of \textit{adult comorbidities}, except for the variable KIDNEY, which is now placed in the third factor together with INMUNO and SMOKE. These two latter variables were not placed in any of the factors with our model due to their very small coefficients (less than $0.10$). Finally, the new second factor contains HOSP and PNEUM with positive coefficients and WOMEN with a negative coefficient. We can say the this new factor combines our first (\textit{covid complications}) and second (\textit{gender}) factors.

\section{Concluding remarks}
\label{sec:conclusion}

We have proposed a new factor model for the analysis of dichotomous variables. The model assumes a Bernoulli distribution for the responses and a Dirichlet distribution for the factors. All model parameters $\Omega$ and $\balpha$ are interpretable. 

Since the inferential procedure is Bayesian and relies on proper prior distributions, the parameters are all estimable. However, the autocorrelation in the MCMC chains for the $\alpha_k$'s cannot be disregarded, so long chains with a large thinning step are required. 

The main code to implement our model, that is, the MCMC procedure, is programmed in Fortran, and it is called from the \cite{r-package:26}. Both the Fortran and  R codes, together with the two datasets as well as the JAGS code, are available as Supplementary Material. 

In future work, we are planning to create a contributed package in R to make this model accessible to a larger community. 

\section*{Acknowledgements}
This work was supported by \textit{Asociaci\'on Mexicana de Cultura, A.C.}

\bibliographystyle{natbib}

\begin{thebibliography}{99}

\bibitem[Bartlett, 1937]{bartlett:37}
Bartlett, M.S. (1937). The statistical conception of mental factors. \textit{British journal of psychology} \textbf{28}, 97--104. 

\bibitem[Beauducel and Hilger, 2015]{beaducel&hilger:15}
Beauducel, A. and Hilger, N. (2015). Extending the debate between Spearman and Wilson 1929: When do single variables optimally reproduce the common part of the observed covariances? \textit{Multivariate Behavioral Research} \textbf{50}, 555--567.

\bibitem[Bonett and Price, 2005]{bonett&price:05}
Bonett, D.G. and Price, R.M. (2005). Inferential methods for the tetrachoric correlation coefficient. \textit{Journal of Educational and Behavioral Statistics} \textbf{30}, 213--225.

\bibitem[Geisser and Eddy, 1979]{geisser&eddy:79}
Geisser, S. and Eddy, W.F. (1979). A predictive approach to model selection. \textit{Journal of the American Statistical Association} \textbf{74}, 153--160. 

\bibitem[Hendrickson and White, 1964]{hendrickson&white:64}
Johnson, R.A. and Wichern, D.W. (2002). \textit{Applied Multivariate Statistical Analysis}. Prentice Hall, New Jersey. 

\bibitem[Johnson and Wichern, 2002]{johnson&wichern:02}
Hendrickson, A.E. and White, P.O. (1964). Promax: A quick method for rotation to oblique simple structure. \textit{The British Journal of Statistical Psychology} \textbf{17}, 65--70. 

\bibitem[Kaiser, 1958]{kaiser:58}
Kaiser, H.F. (1958). The varimax criterion for analytic rotation in factor analysis. \textit{Psychometrika} \textbf{23}, 187--200.

\bibitem[Muth\'en, 1978]{muthen:78}
Muth\'en, B.O. (1978). Contributions to factor analysis of dichotomous variables. \textit{Psychometrika} \textbf{43}, 551--560.

\bibitem[Nieto-Barajas, 2025]{nieto:25}
Nieto-Barajas, L.E. (2025). \textit{Dependence models via hierarchical structures}. Cambridge University Press. 

\bibitem[Plummer, 2023]{plummer:23}
Plummer, M. \textit{rjags: Bayesian Graphical Models using MCMC}, 4rd ed.
R-package, CRAN, 2023.

\bibitem[R-package, 2026]{r-package:26}
R Core Team (2026). \textit{R: A Language and Environment for Statistical Computing}. R Foundation for Statistical Computing, Vienna, Austria.

\bibitem[Roberts and Rosenthal, 2009]{roberts&rosenthal:09} 
Roberts, G.O. and Rosenthal, J.S. (2009). Examples of adaptive MCMC. \textit{Journal of Computational and Graphical Statistics} \textbf{18}, 349--367.

\bibitem[Takane and Leeuw, 1987]{takane&leeuw:87}
Takane and de Leeuw. (1987). On the relationship between item response theory and factor analysis of discretized variables. \textit{Psychometrika} \textbf{52}, 393. 

\bibitem[Tierney, 1994]{tierney:94}
Tierney, L. (1994). Markov chains for exploring posterior distributions. \textit{Annals of Statistics} \textbf{22}, 1701--1762.

\end{thebibliography}

\newpage

\begin{table}
$$\begin{array}{ccc|ccc|ccc} \hline \hline \\[-4mm]
\multicolumn{3}{c|}{\Omega} & \multicolumn{3}{c|}{\widehat\Omega_1} & \multicolumn{3}{c}{\widehat\Omega_2} \\ \hline
0.01 & \boxed{0.75} & 0.13 & 0.14 & 0.02 & \boxed{0.80} & \boxed{0.79} & 0.15 & 0.02 \\ 
0.05 & \boxed{0.96} & 0.21 & 0.29 & 0.05 & \boxed{0.88} & \boxed{0.87} & 0.29 & 0.06 \\
0.24 & 0.39 & \boxed{0.92} & \boxed{0.84} & 0.31 & 0.37 & 0.38 & \boxed{0.83} & 0.32 \\
0.04 & \boxed{0.75} & 0.03 & 0.21 & 0.03 & \boxed{0.64} & \boxed{0.63} & 0.21 & 0.04 \\
\boxed{0.99} & 0.03 & 0.16 & 0.19 & \boxed{0.95} & 0.06 & 0.07 & 0.19 & \boxed{0.94} \\
\boxed{0.88} & 0.12 & 0.02 & 0.03 & \boxed{0.96} & 0.08 & 0.09 & 0.03 & \boxed{0.95} \\
0.42 & \boxed{0.91} & 0.01 & 0.03 & 0.58 & \boxed{0.89} & \boxed{0.88} & 0.03 & 0.58 \\
\hline \hline
\end{array}$$
\caption{Simulated data. Original and two point estimates of the loading matrix.}
\label{tab:sim}
\end{table}

\begin{table}
$$\begin{array}{c|cccccc} \hline \hline
{\rm Data} & {\rm WOMEN} & {\rm HOSP} & {\rm PNEUM} & {\rm ADULT} & {\rm DIABET} & {\rm EPOC} \\ 
{\rm covid} & 0.59 & 0.46 & 0.24 & 0.33 & 0.14 & 0.03 \\ \hline
{\rm Data} & {\rm ASMA} & {\rm INMUNO} & {\rm CARDIO} & {\rm OBESITY} & {\rm KIDNEY} & {\rm SMOKE} \\
{\rm covid} & 0.03 & 0.06 & 0.17 & 0.07 & 0.04 & 0.05 \\
\hline \hline
\end{array}$$
\caption{COVID19 dataset. Proportions for each indicator variable.}
\label{tab:eda}
\end{table}

\begin{table}
$$\begin{array}{c|cc|ccc|cccc} \hline \hline \\[-4mm]
X_j & \multicolumn{2}{c|}{\widehat\Omega_2} & \multicolumn{3}{c|}{\widehat\Omega_3^{varimax}} & \multicolumn{4}{c}{\widehat\Omega_4^{varimax}} \\ \hline
\rm{WOMEN} & \boxed{0.63} & 0.43 & 0.28 & \boxed{0.88} & 0.25 & 0.21 & 0.24 & 0.00 & 1.14 \\
\rm{HOSP} & 0.32 & \boxed{0.97} & \boxed{1.15} & 0.43 & 0.51 & 1.15 & 0.55 & 0.00 & 0.46 \\
\rm{PNEUM} & 0.12 & \boxed{0.69} & \boxed{0.66} & 0.24 & 0.24 & 0.71 & 0.26 & 0.00 & 0.24 \\
\rm{ADULT} & 0.14 & \boxed{0.99} & 0.37 & 0.38 & \boxed{0.87} & 0.35 & 0.87 & 0.01 & 0.45 \\
\rm{DIABET} & 0.00 & \boxed{0.66} & 0.21 & 0.14 & \boxed{0.69} & 0.21 & 0.73 & 0.00 & 0.13 \\
\rm{EPOC} & 0.00 & \boxed{0.13} & 0.04 & 0.03 & \boxed{0.14} & 0.05 & 0.15 & 0.00 & 0.03 \\
\rm{ASMA} & 0.03 & 0.03 & 0.03 & 0.04 & 0.02 & 0.03 & 0.01 & 0.00 & 0.06 \\
\rm{INMUNO} & 0.04 & \boxed{0.12} & 0.09 & 0.07 & 0.07 & 0.09 & 0.07 & 0.00 & 0.09 \\
\rm{CARDIO} & 0.00 & \boxed{0.85} & 0.26 & 0.18 & \boxed{0.88} & 0.27 & 0.91 & 0.00 & 0.17 \\
\rm{OBESITY} & 0.03 & \boxed{0.25} & 0.06 & 0.09 & \boxed{0.24} & 0.06 & 0.25 & 0.00 & 0.11 \\
\rm{KIDNEY} & 0.00 & \boxed{0.20} & 0.06 & 0.04 & \boxed{0.21} & 0.07 & 0.22 & 0.00 & 0.04 \\
\rm{SMOKE} & 0.03 & \boxed{0.10} & 0.04 & 0.06 & 0.08 & 0.04 & 0.09 & 0.00 & 0.07 \\ \hline
\widehat\alpha & 1.16 & 0.32 & 0.30 & 0.57 & 0.22 & 0.27 & 0.19 & 0.30 & 0.30 \\
Q_{0.025} & 0.98 & 0.29 & 0.27 & 0.49 & 0.20 & 0.23 & 0.17 & 0.25 & 0.25 \\
Q_{0.975} & 1.37 & 0.36 & 0.35 & 0.68 & 0.26 & 0.33 & 0.23 & 0.38 & 0.36 \\
\hline \hline
\end{array}$$
\caption{COVID19 data. Point estimates of the loading matrix $\Omega$ for different number of factors. Posterior point estimates and 95\% CI of $\alpha_k$, $k=1,\ldots,q$ (last three rows).}
\label{tab:real}
\end{table}

\begin{table}
$$\begin{array}{c|ccc} \hline \hline \\[-4mm]
X_j & \multicolumn{3}{c}{\widehat\Lambda} \\ \hline
\rm{WOMEN} & -0.01 & \boxed{-0.25} & -0.05 \\
\rm{HOSP}& 0.13 & \boxed{0.96} & 0.25 \\
\rm{PNEUM} & 0.18 & \boxed{0.85} & 0.06 \\
\rm{ADULT} & \boxed{0.81} & 0.07 & 0.16 \\
\rm{DIABET} & \boxed{0.76} & 0.08 & 0.34 \\
\rm{EPOC} & \boxed{0.73} & 0.35 & -0.17 \\
\rm{ASMA} & -0.02 & 0.08 & -0.04 \\
\rm{INMUNO} & 0.01 & 0.01 & \boxed{0.41} \\
\rm{CARDIO} & \boxed{0.71} & 0.08 & 0.47 \\
\rm{OBESITY} & \boxed{0.45} & -0.07 & 0.10 \\
\rm{KIDNEY} & 0.32 & 0.20 & \boxed{0.92} \\
\rm{SMOKE} & 0.13 & 0.00 & \boxed{0.19} \\ \hline
\rm{SS\;load.} & 2.65 & 1.88 & 1.53 \\
\rm{Prop.Var} & 0.22 & 0.16 & 0.13 \\
\rm{Cum.Var} & 0.22 & 0.38 & 0.51 \\
\hline \hline
\end{array}$$
\caption{COVID19 data. Normal factor analysis on tetrachorich correlations. Sum of loadings squares, proportion of variance and cumulative variance (last three rows).}
\label{tab:tetra}
\end{table}

\end{document}